\documentclass[longauth]{aa} 

\usepackage{graphicx, array}
\usepackage[varg]{txfonts}
\usepackage{natbib,twoopt}
\usepackage{xcolor}
\usepackage[breaklinks=true]{hyperref} 
\bibpunct{(}{)}{;}{a}{}{,}             
\makeatletter
  \newcommandtwoopt{\citeads}[3][][]{\href{http://adsabs.harvard.edu/abs/#3}%
    {\def\hyper@linkstart##1##2{}%
     \let\hyper@linkend\@empty\citealp[#1][#2]{#3}}}
  \newcommandtwoopt{\citepads}[3][][]{\href{http://adsabs.harvard.edu/abs/#3}%
    {\def\hyper@linkstart##1##2{}%
     \let\hyper@linkend\@empty\citep[#1][#2]{#3}}}
  \newcommandtwoopt{\citetads}[3][][]{\href{http://adsabs.harvard.edu/abs/#3}%
    {\def\hyper@linkstart##1##2{}%
     \let\hyper@linkend\@empty\citet[#1][#2]{#3}}}
  \newcommandtwoopt{\citeyearads}[3][][]%
    {\href{http://adsabs.harvard.edu/abs/#3}
    {\def\hyper@linkstart##1##2{}%
     \let\hyper@linkend\@empty\citeyear[#1][#2]{#3}}}
\makeatother
\hypersetup{colorlinks=true,linkcolor=blue,citecolor=blue,urlcolor=blue}

\newcommand{\kms}{km~s$^{-1}$}
\newcommand{\ms}{m~s$^{-1}$}
\newcommand{\srv}{$\sigma_{\rm RV}$}
\newcommand{\harps}{{\small HARPS}}
\newcommand{\sophie}{{\small SOPHIE}}
\newcommand{\elodie}{{\small ELODIE}}
\newcommand{\narval}{{\small NARVAL}}
\newcommand{\coralie}{{\small CORALIE}}

\newcommand{\gaia}{\textit{Gaia}}
\newcommand{\RVS}{\textit{RVS}}
\newcommand\hip{\textsc{Hipparcos}}

\newcommand\gdrone{\gaia~DR1}
\newcommand\gdrtwo{\gaia~DR2}
\newcommand\gdrthree{\gaia~DR3}
\newcommand{\grvs}{$G_\mathrm{RVS}$}

\begin{document}

   \title{\gaia\ Data Release 2 }
   \subtitle{The catalogue of radial velocity standard stars
   \thanks{Based on observations made at Observatoire de Haute Provence (CNRS), France, at the Telescope Bernard Lyot (USR5026) operated by the Observatoire Midi-Pyr\'en\'ees, Universit\'e de Toulouse (Paul Sabatier) and CNRS, France, at the Euler telescope operated by Observatoire de Gen\`eve  at La Silla, Chile, and on public data obtained from the ESO Science Archive Facility.}
\thanks{Tables are only available in electronic form at the CDS via anonymous ftp to cdsarc.u-strasbg.fr (?) 
or via http://cdsarc.u-strasbg.fr/viz-bin/qcat?J/}
   }

    \author{
         C.  ~Soubiran                   \inst{\ref{inst:0001}}
\and G.  ~Jasniewicz                 \inst{\ref{inst:0002}}
\and L.  ~Chemin                        \inst{\ref{inst:0003}}
\and C. ~Zurbach                      \inst{\ref{inst:0002}}
\and N. ~Brouillet                  \inst{\ref{inst:0001}}
\and P.  ~Panuzzo                         \inst{\ref{inst:0004}}
\and P.  ~Sartoretti                       \inst{\ref{inst:0004}}
\and D. ~Katz                                  \inst{\ref{inst:0004}}
\and J.-F.  ~Le Campion                  \inst{\ref{inst:0001}}
\and O.  ~Marchal                        \inst{\ref{inst:0004}}
\and D.  ~Hestroffer                       \inst{\ref{inst:0005}}
\and F.  ~Th\'evenin                       \inst{\ref{inst:0006}}
\and F.  ~Crifo                         \inst{\ref{inst:0004}}
\and S.  ~Udry                       \inst{\ref{inst:0007}}
\and M.      ~Cropper                       \inst{\ref{inst:0008}}  
\and G.      ~Seabroke                   \inst{\ref{inst:0008}}
\and Y.       ~Viala                          \inst{\ref{inst:0004}}
\and K.      ~Benson                       \inst{\ref{inst:0008}}  
\and R.       ~Blomme                    \inst{\ref{inst:0009}}
\and A.       ~Jean-Antoine            \inst{\ref{inst:0010}}
\and H.      ~Huckle                      \inst{\ref{inst:0008}}  
\and M.      ~Smith                       \inst{\ref{inst:0008}}
\and S. G.      ~Baker                     \inst{\ref{inst:0008}} 
\and Y.        ~Damerdji                \inst{\ref{inst:0011},\ref{inst:0013}}
\and C.       ~Dolding                  \inst{\ref{inst:0008}}
\and Y.        ~Fr\'{e}mat                \inst{\ref{inst:0009}}\relax
\and E.        ~Gosset                 \inst{\ref{inst:0013},\ref{inst:0014}}\relax
\and A.        ~Guerrier                 \inst{\ref{inst:0010b}}
\and L.P.      ~Guy                           \inst{\ref{inst:0015}}\relax
\and R.        ~Haigron                       \inst{\ref{inst:0004}}\relax
\and K.        ~Jan{\ss}en              \inst{\ref{inst:0016}}\relax
\and G.       ~Plum                      \inst{\ref{inst:0004}}
\and C.       ~Fabre                        \inst{\ref{inst:0010c}}
\and Y.         ~Lasne                      \inst{\ref{inst:0010b}}
\and F.        ~Pailler                       \inst{\ref{inst:0010}}
\and C.       ~Panem                     \inst{\ref{inst:0010}}
\and F.       ~Riclet                       \inst{\ref{inst:0010}}
\and F.       ~Royer                       \inst{\ref{inst:0004}}  
\and G.       ~Tauran                    \inst{\ref{inst:0010b}}
\and T.       ~Zwitter                       \inst{\ref{inst:0017}}\relax 
\and A.       ~Gueguen                 \inst{\ref{inst:0004},\ref{inst:0018}} 
\and C.         ~Turon                        \inst{\ref{inst:0004}} 
  }
  
  \institute{
Laboratoire d'astrophysique de Bordeaux, Univ. Bordeaux, CNRS, B18N, all{\'e}e Geoffroy Saint-Hilaire, F-33615 Pessac, France\relax                                                           
\label{inst:0001}   
\email{caroline.soubiran@u-bordeaux.fr}
\and Laboratoire Univers et Particules de Montpellier, Universit\'{e} Montpellier, CNRS, Place Eug\`{e}ne Bataillon, CC72,  F-34095 Montpellier Cedex 05, France\relax                                               
 \label{inst:0002}
\and Unidad de Astronom\'ia, Fac. Cs. B\'asicas, Universidad de Antofagasta, Avda. U. de Antofagasta 02800, Antofagasta, Chile\relax                                               
 \label{inst:0003}
 \and GEPI, Observatoire de Paris, Universit{\'e} PSL, CNRS,  5 Place Jules Janssen,  F-92190 Meudon, France\relax                                                                                             
 \label{inst:0004}
\and IMCCE, Observatoire de Paris, PSL Research University, CNRS, Sorbonne Universit{\'e}, Univ. Lille, 77 av. Denfert-Rochereau,  F-75014 Paris, France\relax                                                       
\label{inst:0005}
 \and Universit{\'e} C{\^o}te d'Azur, Observatoire de la C{\^o}te d'Azur, CNRS, Laboratoire Lagrange, Bd de l'Observatoire, CS 34229,  F-06304 Nice cedex 4, France\relax                                                       
 \label{inst:0006}
\and Observatoire de Gen\`eve, Universit\'e de Gen\`eve, 51 Ch. des Maillettes, CH-1290 Sauverny, Switzerland\relax                                                       
 \label{inst:0007}
  \and Mullard Space Science Laboratory, University College London, Holmbury St Mary, Dorking, Surrey RH5 6NT, United Kingdom\relax                                                                            \label{inst:0008}
\and Royal Observatory of Belgium, Ringlaan 3, B-1180 Brussels, Belgium\relax                                                                                                                                 
 \label{inst:0009}
\and CNES Centre Spatial de Toulouse, 18 avenue Edouard Belin,  F-31401 Toulouse Cedex 9, France\relax                                                                                                          
\label{inst:0010}
\and CRAAG - Centre de Recherche en Astronomie, Astrophysique et G\'{e}ophysique, Route de l'Observatoire Bp 63 Bouzareah 16340, Algiers, Algeria\relax                                                      
 \label{inst:0011}
 \and Institut d'Astrophysique et de G\'{e}ophysique, Universit\'{e} de Li\`{e}ge, 19c, All\'{e}e du 6 Ao\^{u}t, B-4000 Li\`{e}ge, Belgium\relax   
\label{inst:0013}
\and F.R.S.-FNRS, Rue d'Egmont 5, B-1000 Brussels, Belgium\relax                                                                                                                                              
 \label{inst:0014}
\and Thales Services, 290 All\'ee du Lac,  F-31670 Lab\`ege, France\relax                                                                                                          
\label{inst:0010b}
\and Department of Astronomy, University of Geneva, Chemin d'Ecogia 16, CH-1290 Versoix, Switzerland\relax                                                                                                   
\label{inst:0015}
 \and Leibniz Institute for Astrophysics Potsdam (AIP), An der Sternwarte 16, D-14482 Potsdam, Germany\relax                                                                                                    
 \label{inst:0016}
\and ATOS for CNES Centre Spatial de Toulouse, 18 avenue Edouard Belin,  F-31401 Toulouse Cedex 9, France\relax 
\label{inst:0010c}
\and Faculty of Mathematics and Physics, University of Ljubljana, Jadranska ulica 19, SI-1000 Ljubljana, Slovenia\relax                                                                                        
\label{inst:0017}
\and Max Planck Institute for Extraterrestrial Physics, High Energy Group, Gie{\ss}enbachstra{\ss}e, D-85741 Garching, Germany\relax
\label{inst:0018}
}

 \date{Received \today, accepted }

 
  \abstract
   {}
   {The Radial Velocity Spectrometer (RVS) on board the ESA satellite mission  \gaia\  has no calibration device. Therefore, the radial velocity zero point needs to be calibrated with stars that are proved to be stable  at a level of 300 \ms\ during the \gaia\  observations. }
   {We compiled a dataset of $\sim$71\,000 radial velocity measurements from five high-resolution spectrographs. A catalogue of 4\,813 stars was built  by combining these individual measurements. The zero point was established using asteroids.}
   {The resulting catalogue has  seven observations per star on average on a typical time baseline of six years, with a median standard deviation of 15 \ms.  A subset of the most stable stars fulfilling the \RVS\ requirements was used to establish the radial
velocity zero point provided in \gaia\  Data Release 2. The stars that were not used for calibration are used to validate the \RVS\ data.}
   {}


   \maketitle
%

\section{Introduction}

   After a successful launch on 19 December 2013 and after several months of commissioning activities, the \gaia\ satellite started its scientific observations in July 2014 \citep{gdr1a}. \gaia\  is
   currently scanning the sky continuously and  observes all objects of magnitude  $2 \lesssim G \le 20.7$.
   In addition to the position, parallax, and proper motion that
is measured for more than one billion stars, the radial velocity (RV) is also determined for 
   the $\sim$150 million brightest stars (\grvs $\lesssim$16.2)  with the Radial Velocity Spectrometer \citep[\RVS,][]{kat04,DR2-DPACP-46}.  
   The first \RVS\ measurements are published in the \gaia\  Data Release 2 \citep[DR2,][]{DR2-DPACP-47}, providing an unprecedented 6D map of the Milky Way \citep{DR2-DPACP-33}.  
   The \RVS\ is an integral-field spectrograph with resolving power of $\sim$ 11\,500 covering the near-infra-red wavelength range at 845--872 nm. The \RVS\ will record 40 epochs on average per source 
   during the five years of the nominal mission. The RV precision is expected to be 1 \kms\ for GK type stars down to $G$ = 12-13. The summary of the contents and survey properties of \gaia\  DR2 are presented in \cite{DR2-DPACP-36}. 
   The latest news about \gaia\  can be found on the ESA website (http://www.cosmos.esa.int/web/gaia). \\
   
   The \RVS\ has no calibration device. As explained in \cite{DR2-DPACP-47},  the wavelength calibration is based on all bright, well-behaved, and stable FGK-type stars observed by the \RVS. The zero point (ZP hereafter)
   of the RVs needs to be fixed with RV standard stars (RV-STDs),
however, that are known in advance and proved to be stable at a level of 300 \ms\ during the \gaia\  observations. When we started to search for suitable RV-STDs in 2006, no
   catalogue existed that would have fulfilled the \RVS\ requirements in terms of number of stars, magnitude range, sky coverage, and precision. A first list of 1420 RV-STD candidates was established \citep{cri07,cri10} that are
   all part of the \hip\ catalogue and selected in three sources of RVs: ``Radial velocities of 889 late-type stars'' \citep{nid02},  ``Radial velocities for 6691 K and M giants'' \citep{fam05}, and 
   ``The Geneva-Copenhagen Survey of the Solar neighbourhood'' \citep{nor04}, complemented with IAU standards \citep{udr99}. We observed the selected stars between 2006 and 2012 with high-resolution spectrometers. This observing campaign was part of a more general task of the \gaia\ Data Processing and Analysis Consortium (DPAC), which was to acquire all necessary ground-based observations for \gaia\ (GBOG). Public archives were queried to complement our observations. The resulting pre-launch version of the catalogue of RV-STDs for \gaia,\  including 10\,214 RV measurements for the 1\,420 stars presented in \cite{sou13}, is now superseded by the version presented here.\\
 
 The number of RV-STDs needed for the ZP calibration has been re-evaluated in December 2013. The first estimate of  $\sim$1000 necessary stars in 2006 was based on the fact that \gaia\  should observe one RV-STD per hour. Later studies showed that the \RVS\ calibration needs at least twice as many RV-STDs. Additional candidates were searched for in the public archives of high-resolution spectrographs, taking advantage of the  large number of FGK-type stars that have been followed-up in exoplanet programmes. New observations were also obtained for the GBOG programme in order to verify the stability of the RV-STD candidates during the \gaia\  operations. \\
   
 The compilation of our own observations and of the measurements retrieved from public archives results in 71\,225 RVs of 4\,813 stars and is described in Sect.~\ref{s:compil}. 
 Great care has been taken to place the catalogue at a ZP defined by asteroids (Sect.~\ref{s:zp}). Section \ref{s:std} recalls the requirements for the calibration of the \RVS\ ZP and presents the 2\,712  RV-STDs 
 that fulfill them. The masks used in the pipelines for the RV determination are obviously very important within the context of the ZP 
 and are consequently discussed in detail. We compare our catalogue to other ground-based catalogues in Sect.~\ref{s:comp} and to the \RVS\ catalogue in Sect.~\ref{s:rvs}. The catalogue is made of two tables that are only available as online material, one with the basic information on the stars, mean ground-based RVs, \gaia\  RVs when available, and RV errors, and the other table with the individual original measurements.

\section{Building the compilation}
\label{s:compil}
For the pre-launch version of the catalogue  \citep{sou13}, we started from the preselected list of good RV-STD candidates established by \cite{cri10}, and we gathered observations a posteriori to verify their stability. In order to enlarge the catalogue,  we did the inverse. We searched in public archives of high-resolution spectrographs  for well observed stars showing stable RV measurements, and we verified afterwards that they fulfilled the \RVS\ requirements for the ZP calibration. This started by the compilation of a large catalogue of RV measurements, which, for consistency with our previous work, come from four spectrographs that specialized in high-precision RVs, the \elodie\ spectrograph \citep{bar96} and the \sophie\ spectrograph \citep{per08}  on the 1.93 m telescope at Observatoire de Haute-Provence (OHP), the \coralie\ spectrograph on the Euler telescope at La Silla Observatory, and the High Accuracy Radial velocity Planet Searcher (\harps) at the ESO La Silla 3.6m telescope \citep{que01,may03}. These four instruments have similar reduction pipelines.  We also used the spectropolarimeter \narval\ \citep{aur03} mounted on the T\'elescope Bernard Lyot at Pic du Midi Observatory in order to obtain spectra in the \RVS\ range  for the preparation of the \RVS\ commissioning \citep{cri12}. 
 A short description of these instruments, their resolution, and spectral range can be found in \cite{sou13}. In all cases, barycentric RVs were determined by cross-correlation of  the spectra with numerical masks that were computed with the same technique \citep{bar96}. \\

The \elodie, \sophie,\ and \harps\ archives provide different types of reduced data. We only used the information recorded in the cross-correlation files that resulted from the automated processing pipelines running at the telescope. The query of the \elodie\ archive \citep{mou04} provided RVs for 26\,315 spectra. The latest query of the  \sophie\ archive was performed in September 2016. At this time, the archive contained nearly 100\,000 spectra, of which about half have a usable RV determination. In contrast to the \sophie\ archive, the \harps\ ESO science archive cannot directly be mined with massive queries. We instead used the AMBRE-HARPS catalogue by \cite{dep14}, which provides parameters for 126\,688 scientific spectra, including RV measurements from the ESO pipeline for 95\% of these spectra. An additional query of the ESO HARPS archive was performed in May 2016 on a limited number of stars to retrieve the most recent observations.\\

In addition to these archive measurements, we conducted a large follow-up programme on \sophie\ between 2015 and 2017 in the DPAC-GBOG frame, in order to obtain new observations of a selection of RV-STD candidates lacking recent measurements. The southern RV-STD candidates of the initial list \citep{cri10} have also been re-observed with \coralie\ since 2013. We also updated the \narval\ dataset with observations of several stars that were not part of the initial list. \\

After we gathered all the RV measurements and properly sorted and selected them, the stars needed to be correctly identified with a rule that allowed us to recognize multiple observations of the same star under different names. Only stars that could be unambiguously identified with a Hipparcos, Tycho-2, or Two Micron All Sky Survey (2MASS) number were kept. This constraint ensures that the cross-match with Simbad or other catalogues and with \gaia\ can be made safely in order to retrieve relevant information on the stars.   \\

\begin{figure}[ht!]
\begin{center}
 \includegraphics[width=\columnwidth]{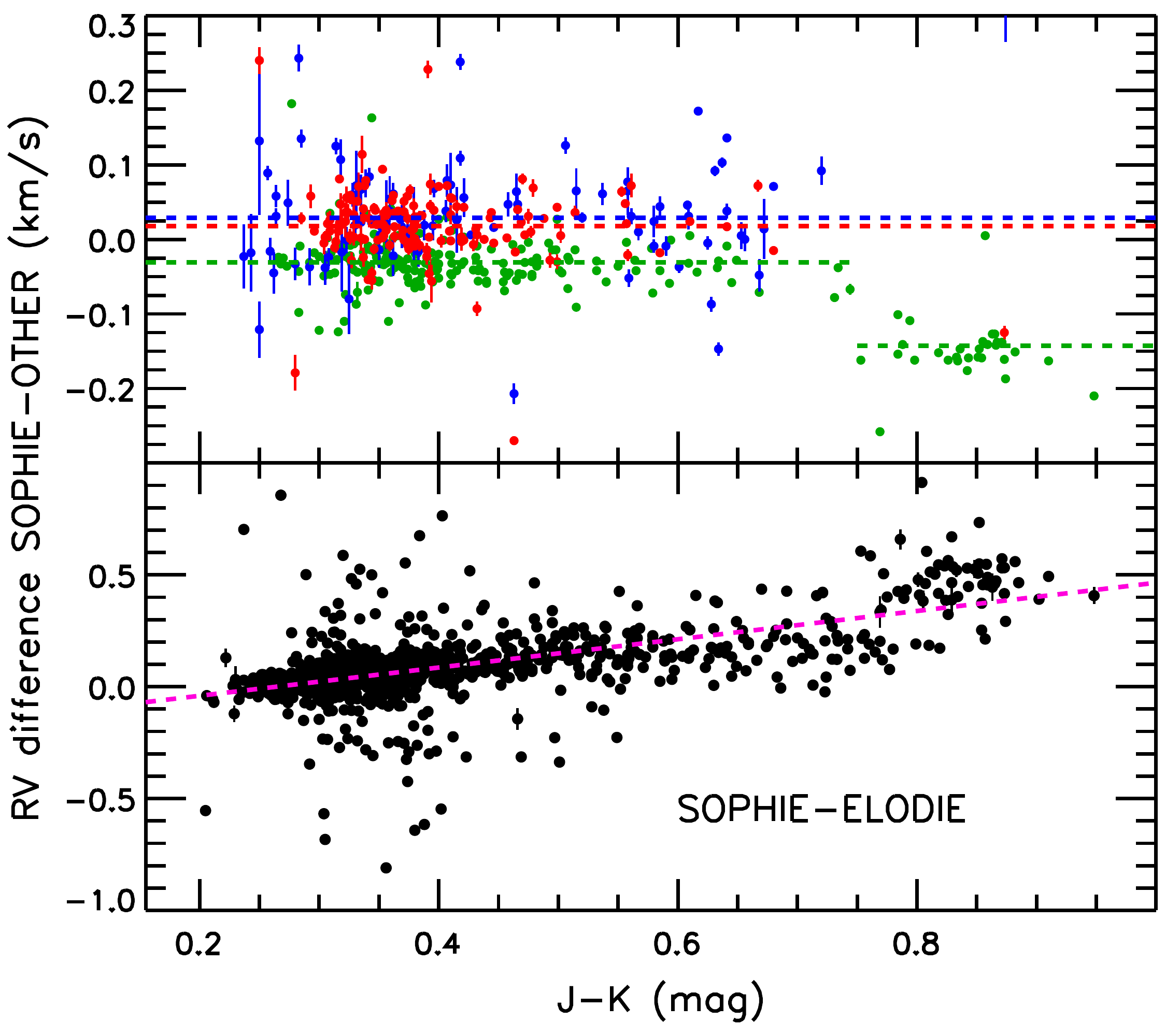}      
 \caption{RV difference between \sophie\ and the other instruments for common stars as a function of the 2MASS colour $J-K$. Top: \coralie\ in red, \narval\ in blue, and \harps\ in green, with the corresponding offsets. Bottom: linear trend with colour of
the RV difference between \sophie\ and \elodie.}
\label{f:so_elo}
\end{center}
\end{figure}

The first step to combine the RV measurements from the different instruments is to express all the RVs in a common system, that
of \sophie. For this, the offsets between \sophie\  and the other instruments were determined. Then, following \cite{sou13}, the mean RV of each star was computed with a weighting scheme based on the uncertainty of the individual measurements. In order to compute the offsets, we derived the weighted  mean of the RV for each instrument. We kept stars whose uncertainties of the mean RV were lower than 100 \ms\ and differed by less than 0.8 \kms. Possible trends with colour were investigated using the 2MASS $J-K$ colour (Fig. \ref{f:so_elo}). The offset \sophie\ $-$ \coralie\ is found to be $18 \pm 3$ \ms based on 147 stars in common, and the offset \sophie\ $-$ \narval\ is found to be $29 \pm 6$ \ms based on 88 stars. These two values agree within the error bar with our previous determination \citep{sou13}. For \elodie\ and \harps,\ the offsets depend on colour, as seen in Fig. \ref{f:so_elo}. Based on 1164 stars in common, the trend \sophie\ $-$ \elodie\ is found to be $0.632(J-K)-0.167 \pm 0.0035$ \kms. For \harps\ the offset is  $-31 \pm 3$ \ms\ for 189 stars with $J-K \leq 0.75$, and $-143 \pm 6$ \ms\ for 28 stars with $J-K > 0.75$. The two colour regimes are clearly seen in Fig. \ref{f:so_elo} and likely result from the use of masks of different spectral types that have different ZPs (see next section). The information about the mask used in the correlation is not available in AMBRE-HARPS, so that a correction depending on the mask instead of colour was not possible. For stars with $J-K \leq 0.75$, the offset is significantly larger than our previous determination ($-17$ \ms), and also larger than the offset determined by \cite{pas12} ($-12$ \ms), possibly because of the mixture of different masks that were used in this colour range. \\

\begin{figure}[ht!]
\begin{center}
 \includegraphics[width=\columnwidth]{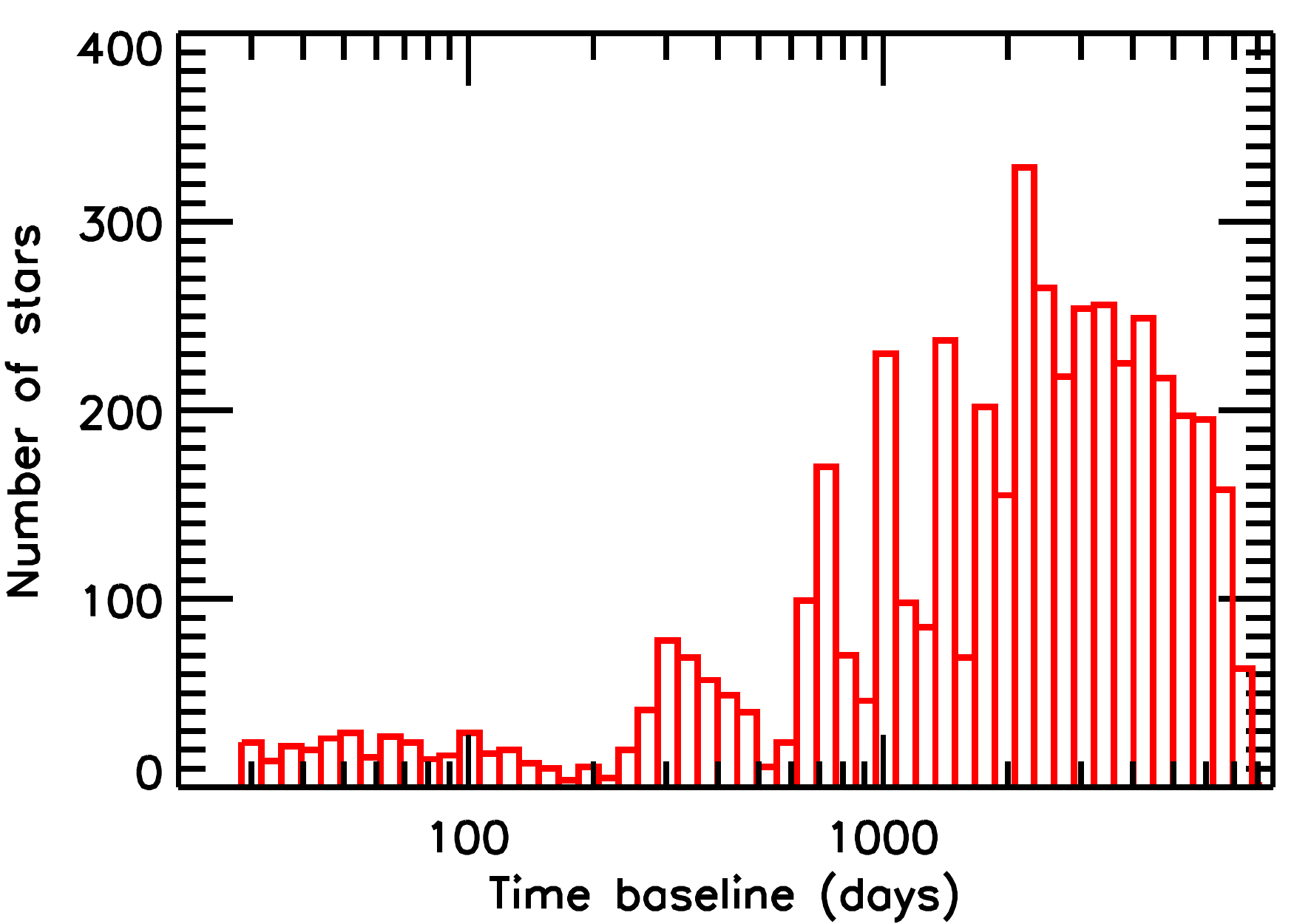}      
 \caption{Time baseline histogram of the observations of each star.}
\label{f:histo_base_t}
\end{center}
\end{figure}

The final version of the catalogue is restricted to stars with at least two observations separated by 30 days that have a weighted standard deviation lower than 0.3 \kms. These thresholds have been adopted after several tests as a compromise between a large number of stars that we wished to keep and the reliability and stability of their RV.
Table~\ref{t:nb_mes} summarizes the information about the RV measurements per instrument. In total, there are 71\,225 RV measurements for 4\,813 stars, 
with a median number of seven observations per star. Figure~\ref{f:histo_base_t} shows the time baseline distribution of the observations per star. The median baseline is 2\,307 days.  \\

\begin{table}[h]
  \centering 
  \caption{Number of RV measurements and stars per instrument, median uncertainty of individual RV measurements $\overline\epsilon,$ and their date range. }
  \label{t:nb_mes}
\begin{tabular}{l|r|r|r|c}
\hline
             & $N_{\rm RV}$ &  $N_{\rm star}$& $\overline \epsilon$  & Date  \\
            &                           &                             &  \ms & range \\
\hline
  \sophie & 26\,680     & 2\,643  & 1.8    & 2006 - 2017 \\
 \harps &   28\,183  &  1\,852 &0.7   & 2003 - 2015 \\
 \elodie & 12\,770    & 1\,795 &5.0  & 1995 - 2006 \\
  \coralie &    3\,293 &  750 &6.6 & 1999 - 2017  \\
 \narval &   299  & 190 &18.8 & 2007 - 2012 \\
\hline
TOTAL & 71\,225 & 4\,813 &1.6 & 1995 - 2017 \\
\hline
\end{tabular}
\end{table}


\begin{figure*}[h!]
\begin{center}
\includegraphics[width=0.5\textwidth]{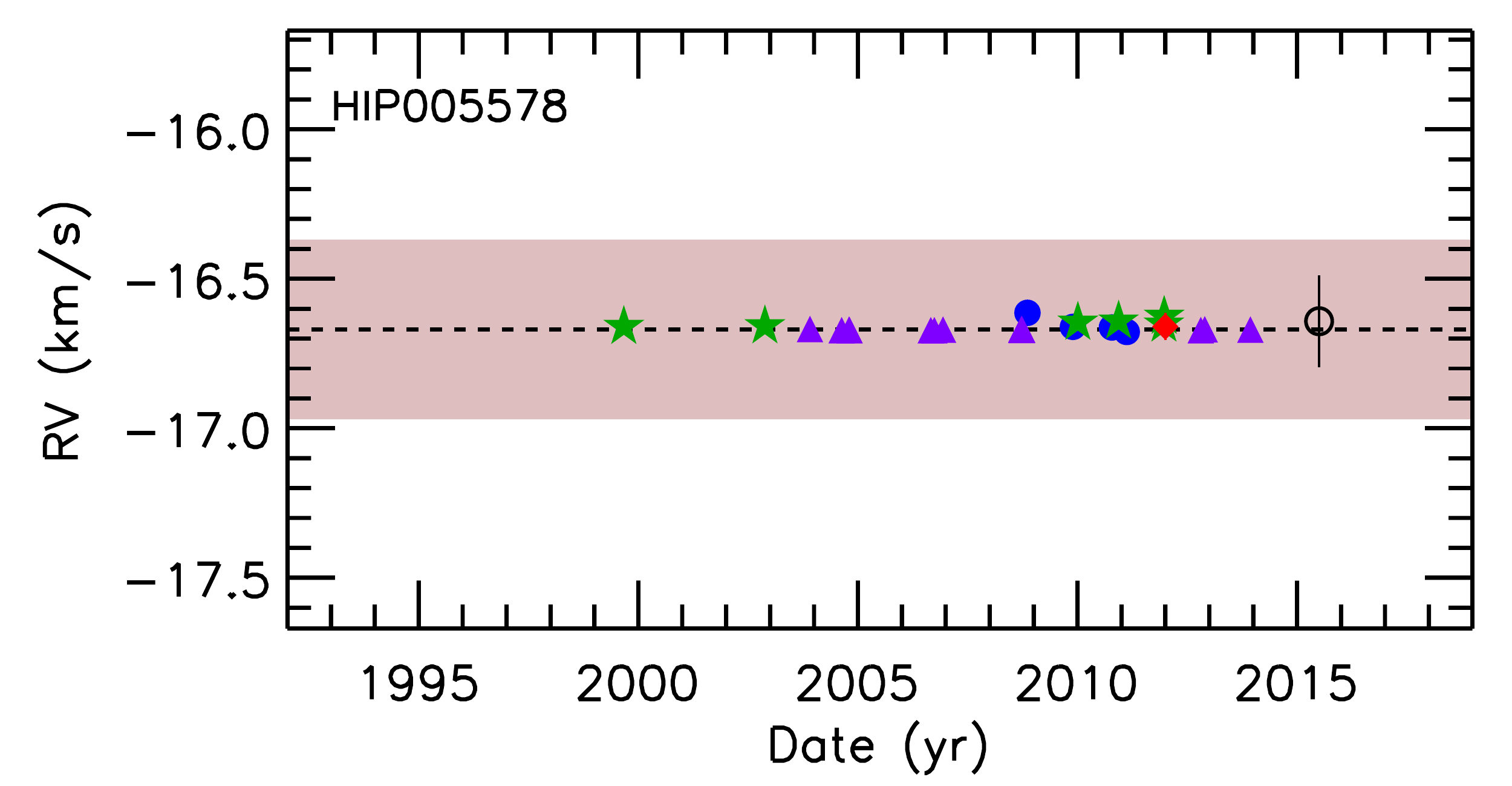}\includegraphics[width=0.5\textwidth]{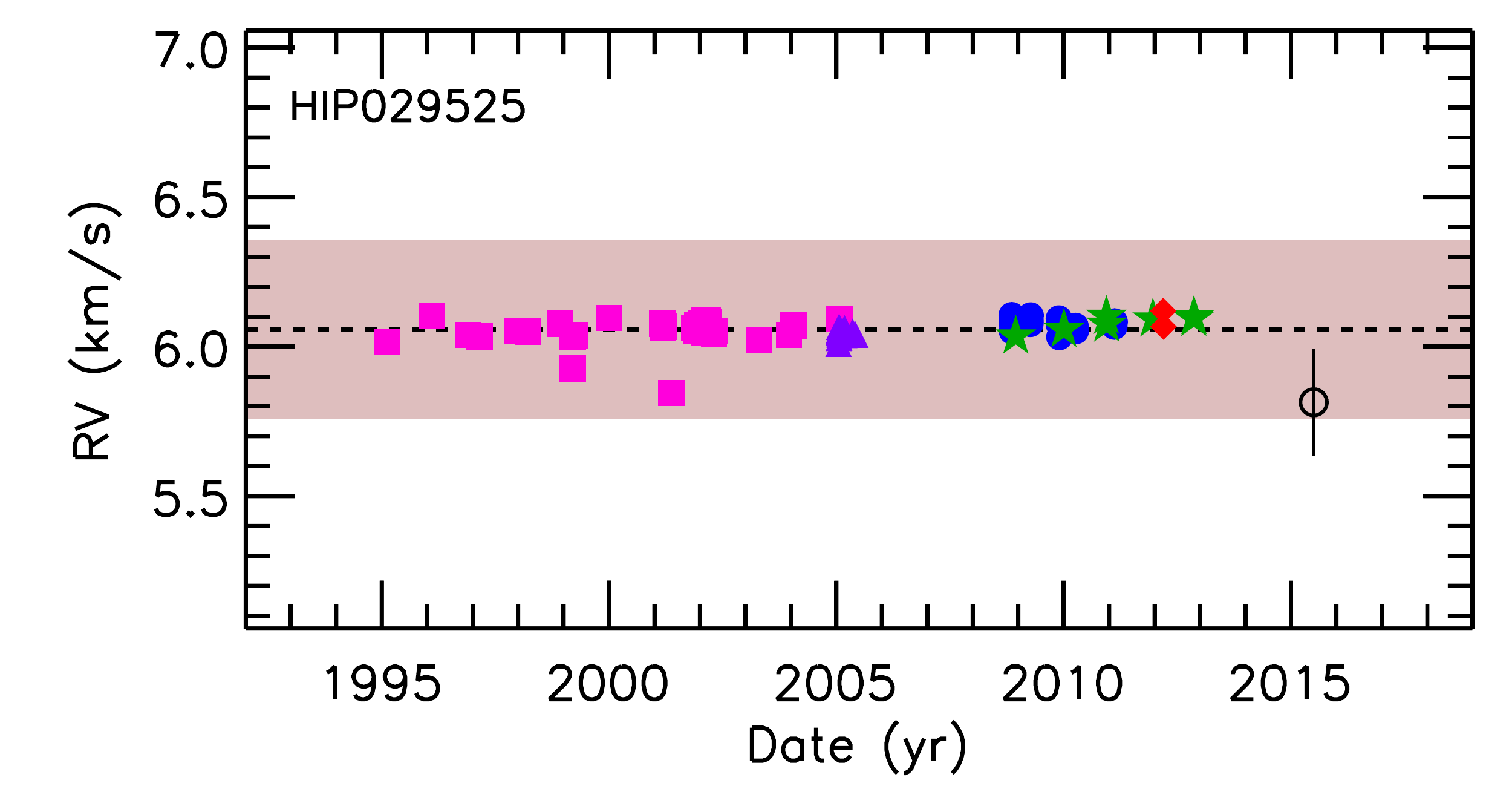}      
 \caption{RV measurements for two of the long-term stable stars of the catalogue. The RV axis is centred on the mean RV of the star and spans 2 \kms. The shaded area represents $\pm$300 \ms. Blue dots represent \sophie\ measurements, green stars \coralie, pink squares \elodie, purple triangles \harps, red diamonds \narval, and open black circles \RVS. The error bars are too small to be visible, except for the \RVS.}
\label{f:stable_stars}
\end{center}
\end{figure*}

Some stars are remarkable for their stability, which is well established by a significant number of observations over a long time baseline. For instance, 35 stars have more than 100 observations over ten years. Two of these stars that were observed with several spectrographs are shown in Fig.~\ref{f:stable_stars}  in order to illustrate the good alignment of the RV measurements from the different instruments. 

\section{Radial velocity zero point}
\label{s:zp}
The determination of RVs by spectroscopy is affected by convective motions and other physical effects that occur in stellar atmospheres  \citep[e.g.][]{dra99}. Considerable efforts have been made in recent years to  evaluate these effects, which strongly depend on the properties of the stars.  The effects of the convective shift on the wavelength of all the spectral lines observed in all
kinds of late-type stars  will be taken into account by using synthetic spectra computed with
3D hydrodynamical models \citep{bon17,chi18}.
The determination of RVs is also affected by the gravitational redshift that results from the deformation of the light path in the potential of the observed star. The gravitational redshift of the Sun measured by \cite{lop00} is 627 \ms, less than 1 \ms\ different from Einstein's prediction. In theory, the gravitational redshift is computable for all stars, but in practice, the value
cannot be obtained directly because the mass and radius are too uncertain for the majority of the stars. The gravitational and  convective shifts, which can reach several hundreds \ms, prevent us from knowing the "true" velocity of a star along the line of sight.  The fundamental definition of RV is fully described by \cite{lin03}, who emphasized the limitations of spectroscopic measurements by gravitational shift, convective shift, other astrophysical effects (stellar rotation, stellar activity, granulation, etc.),  
low-mass companions (stars and exoplanets), and instrumental effects.  \\

In all pipelines attached to the spectrographs we used, the RVs have been obtained by cross-correlating the observed spectra with stellar numerical masks. 
The ways in which the masks were built can differ from one spectrograph to the next and can even differ within a given spectrograph. 
Consequently, the ZP  of an instrument is not unique and depends on the masks. One example is the \sophie\ spectrograph: 
the G2 mask is built from an observed solar spectrum (C. Lovis, private communication), whilst the other masks are adapted from the G2 mask with the help of synthetic spectra of other spectral type.
This fact is of great importance because the gravitational and convective shifts mentioned above are automatically corrected for the solar value in the first 
case, given that cross-correlation is made between two observed spectra, but not necessarily in the second case. In other words, velocities obtained with the SOPHIE G2 
mask are quasi-kinematic RVs for stars similar to the Sun. We  verified that the use of different masks does not alter the ZP for the FGK stars. The situation is more complex for M-type cool dwarfs, whose spectra contain many molecular bands that alter the RV precision. For these stars, we observed RV discrepancies of about 0.4 \kms\ between instruments and between M masks  for a given instrument. Systematic errors of this level in the RVs of M dwarfs have previously been quoted by \citet{mar87}.\\

Another way to adress the ZP problem is to use asteroids, whose kinematic RVs are accurately known from the dynamics of the solar system.
The orbits and motion of the brightest asteroids can indeed be predicted with a precision at the \ms\ level and yield absolute RV values. Hence, in addition to 
regular observations of stars, several asteroids were included in our observing programme, as mentioned in \cite{sou13}. Asteroids have reflectance curves that vary in the continuum 
with the wavelength, but they basically reflect the light and the absorption lines of the Sun. Thus the difference of the RV of an asteroid measured spectroscopically and 
computed via  ephemerides provides a first-order correction of the gravitational and convective shifts for stars similar to the Sun. This method has originally been proposed for 
the calibration of the \RVS\ but was eventually not adopted because
of the faint magnitude of these objects in general and their non-uniform distribution on the sky; they are mostly 
concentrated along the ecliptic. However, the method is very well adapted for ground-based observations \citep[e.g.][]{nid02, zwi07}.\\

\begin{figure}[ht!]
\begin{center}
 \includegraphics[width=\columnwidth]{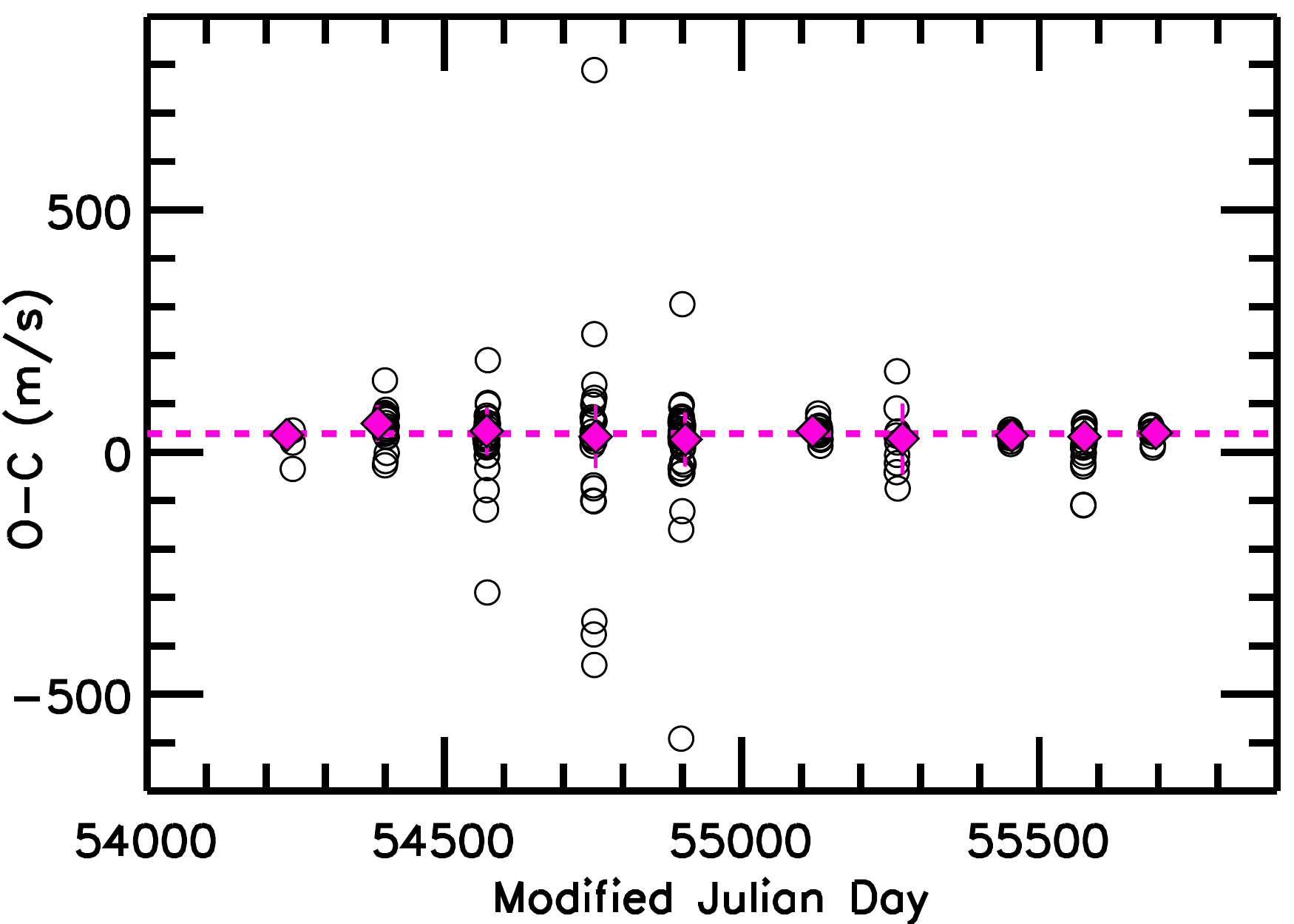}      
\caption{RV $(O-C)$ of asteroids measured at different epochs from 2007 to 2011 as a function of time. Some runs have produced better quality measurements and show a lower dispersion. The red diamonds and error bars represent the weighted mean per run and its uncertainty (values reported in Table~\ref{t:astero}). The horizontal line marks the offset of 38 \ms\ applied to the catalogue to set its ZP on the asteroid scale.}
\label{f:ast}
\end{center}
\end{figure}

During our observations with \sophie,\ we regularly observed asteroids. In total, 81 different asteroids were observed in 
179 exposures during ten observing runs between 2007 and 2011. The number of different asteroids was adopted in order to minimize possible systematic effects arising from their shape, 
rotation, or binarity. The regular observations during five years allowed us to track possible temporal variations. Observations were calibrated and reduced as for the stars
with the same pipeline, yielding the observed instrumental RV. The corresponding kinematical RV was computed using the on-line webservice of ephemerides at IMCCE Miriade\footnote{\url{http://vo.obspm.fr}}. 
The calculation takes into account the motion of the observer with respect to the barycentre of the solar system (BERV), the motion of the asteroid relative to the Sun (giving rise to a change in 
frequency of the reflected light), and last, the motion of the asteroid relative to the observer. The resulting observed minus computed $(O-C)$ RVs are shown in Fig.~\ref{f:ast} as a 
function of time. Some runs show a larger dispersion than others, possibly reflecting the observing conditions. The weighted mean of $(O-C)$ was derived for each run (Table~\ref{t:astero}), 
the weights being defined with the standard uncertainties of the individual measurements. The outliers, measurements at more than 200 \ms\ from the mean, were not included in the calculation. 
The combination of the ten runs gives a weighted average offset of $38\pm5$ \ms. 
This value is low, as expected in the light of the discussion above on the building of the SOPHIE masks.
This offset of 38 \ms\ was applied to our catalogue to set its absolute ZP. However, the version of the catalogue used for 
the calibration and validation of \gdrtwo\ was not yet set to this ZP, which will be effective for \gdrthree. \\

\begin{table}[h]
  \centering 
  \caption{Summary of RV measurements of asteroids observed during ten runs from 2007 to 2011 with \sophie. For each run, the table gives the date as year-month, 
  the number of valid measurements $N_{\rm obs}$, the $(O-C)$ weighted mean, its uncertainty $U_{\rm O-C}$ and standard deviation $\sigma_{\rm O-C}$, and the number of rejected outliers $N_o$.}
  \label{t:astero}
\begin{tabular}{lcrrrrr}
\hline
Run & Date & $N_{\rm obs}$ & $\overline{O-C}$ & $U_{\rm O-C}$ & $\sigma_{\rm O-C}$ & $N_o$ \\
        &         &                              &  \ms                      &   \ms                  &        \ms                      &         \\
\hline
 RUN1  & 2007-05  &  3 &  35.9 &  12.7 &  21.9 & 0\\
 RUN2  & 2007-10  & 25 &  59.3 &   7.6 &  37.9 & 0\\
 RUN3  & 2008-04  & 28 &  44.0 &   9.1 &  48.3 & 1\\
 RUN4  & 2008-10  & 22 &  32.2 &  13.9 &  65.2 & 5\\
 RUN5  & 2009-03  & 33 &  26.3 &   9.8 &  56.2 & 2\\
 RUN6  & 2009-10  & 16 &  44.0 &   4.1 &  16.3 & 0\\
 RUN7  & 2010-03  &  9 &  28.1 &  24.1 &  72.4 & 0\\
 RUN8  & 2010-09  & 10 &  34.4 &   2.7 &   8.5 & 0\\
 RUN9  & 2011-01  & 17 &  31.8 &   7.9 &  32.5 & 0\\
 RUN10 & 2011-05  &  8 &  39.9 &   5.3 &  15.0 & 0\\
 \hline
\end{tabular}
\end{table}

Stars different from the Sun are expected to be differently affected by the gravitational and convective shifts. For instance, the gravitational redshift is 
expected to be low for giant stars. In order to quantify the effects of convective shifts, we checked with \sophie\ data whether stars analysed with the G2 mask and 
another mask (supposedly not corrected of convective shifts) exhibit significant differences in the computed RVs. We considered the 129 stars of our dataset measured with both the G2 and K5 masks. 
The mean difference (K5 $-$ G2) is 34 $\pm$ 7 \ms. For the 20 stars measured with both the G2 and F0 masks,  the mean difference (F0 $-$ G2) is  40 $\pm$ 17 \ms. 
The situation is more complex for cool dwarfs of spectral types M4 and M5. The \sophie\ archive contains many of these stars because they are of interest for exoplanet searches, and 
consequently, many of them are in our catalogue. For these cool dwarfs we see a colour effect when we compared the \sophie\ measurements with \elodie\ and \harps\ and when we compared our
mean catalogue to other catalogues (see next sections). In addition, a systematic difference of 400 \ms\ had to be corrected for 417 \sophie\ correlations that were obtained with an M4 mask, which
was used from 2006 to 2008. 
The M5 mask was still the most frequently used at that time, and it was the only mask since 2008 for very cool dwarfs. A colour trend can be due to different ways in which the templates and masks are build, 
with possibly different synthetic grids, model atmospheres, and line lists. Systematic errors of this level in the RV of M dwarfs have previously been quoted by \cite{mar87}. \\

\section{Radial velocity standards for the RVS}
\label{s:std}

Our RV catalogue includes two categories of stars: the RV-STDs that fulfill the \RVS\ requirements for the ZP calibration, and the other stars that are used for validation of the \RVS\  \citep{DR2-DPACP-54}. In order to select the calibration stars, $B,V,$ and 2MASS magnitudes, spectral types, and object types were retrieved from Simbad.  We also used the absolute magnitudes and binarity information from the XHIP catalogue \citep{xhip}. Additional information was taken from \gdrone\ \citep{gdr1b} and the fourth U.S. Naval Observatory CCD Astrograph Catalogue  \citep[UCAC4, ][]{ucac4}. \\

To become an RV-STD for the \RVS\ ZP calibration, a star must verify  the following requirements, established in 2006:
\begin{itemize}
\item $V_{\rm mag} \leq 11 $;
\item FGK type stars : the spectral types are from Simbad or XHIP. We also applied the colour selection $0.35 \leq B-V \leq 1.40$. Then we individually examined the bluest and reddest stars to eliminate A or M stars. To ensure that we took out as many M dwarfs as possible, we rejected all stars with absolute magnitude $M_V > 8$ in XHIP. It is important to avoid M stars among RV-STDs because of the molecular bands in the \RVS\ range that make their RV determination uncertain;
\item no neighbouring star  with $\Delta I_{\rm mag} < 4$ or brighter within 20\arcsec. This was done with both \gdrone\ and the UCAC4 catalogue. This criterion was relaxed with respect to the initial criterion  \citep[$\Delta I_{\rm mag} < 4$ within 80\arcsec,][]{cri10} because the \RVS\ pipeline is able to reject the transits where a spectrum is polluted by another one;
\item RV stability better than 300 \ms\ over 300 days at least. First we eliminated the binaries as well as possible that were
 identified as such in Simbad or XHIP . Then we considered the stars with a standard deviation, \srv, lower than 100 \ms\ (the level of stability is defined by 3\srv).
\end{itemize}

These requirements were fulfilled by 2\,712 RV-STDs. A slight difference arises between the list of RV-STDs used for \gdrtwo\ and the list that will be used to produce the data for DR3. One reason is that our 2017 observations on \sophie\ were not yet integrated when the \RVS\ processing started for \gdrtwo. Some RV-STD candidates were found to exceed the 100 \ms\ limit in \srv\ with the recent observations and will no longer be used for calibrations. The 26 rejected stars are flagged in the catalogue. The second reason is that the astrometry of some RV-STDs was uncertain at the beginning of the \RVS\ processing, and they
were  excluded from the initial calibration that produced the RVs for \gdrtwo. The corresponding 66 stars were re-integrated in the current calibration set and are also flagged in the catalogue.

\begin{figure}[ht!]
\begin{center}
\includegraphics[width=\columnwidth]{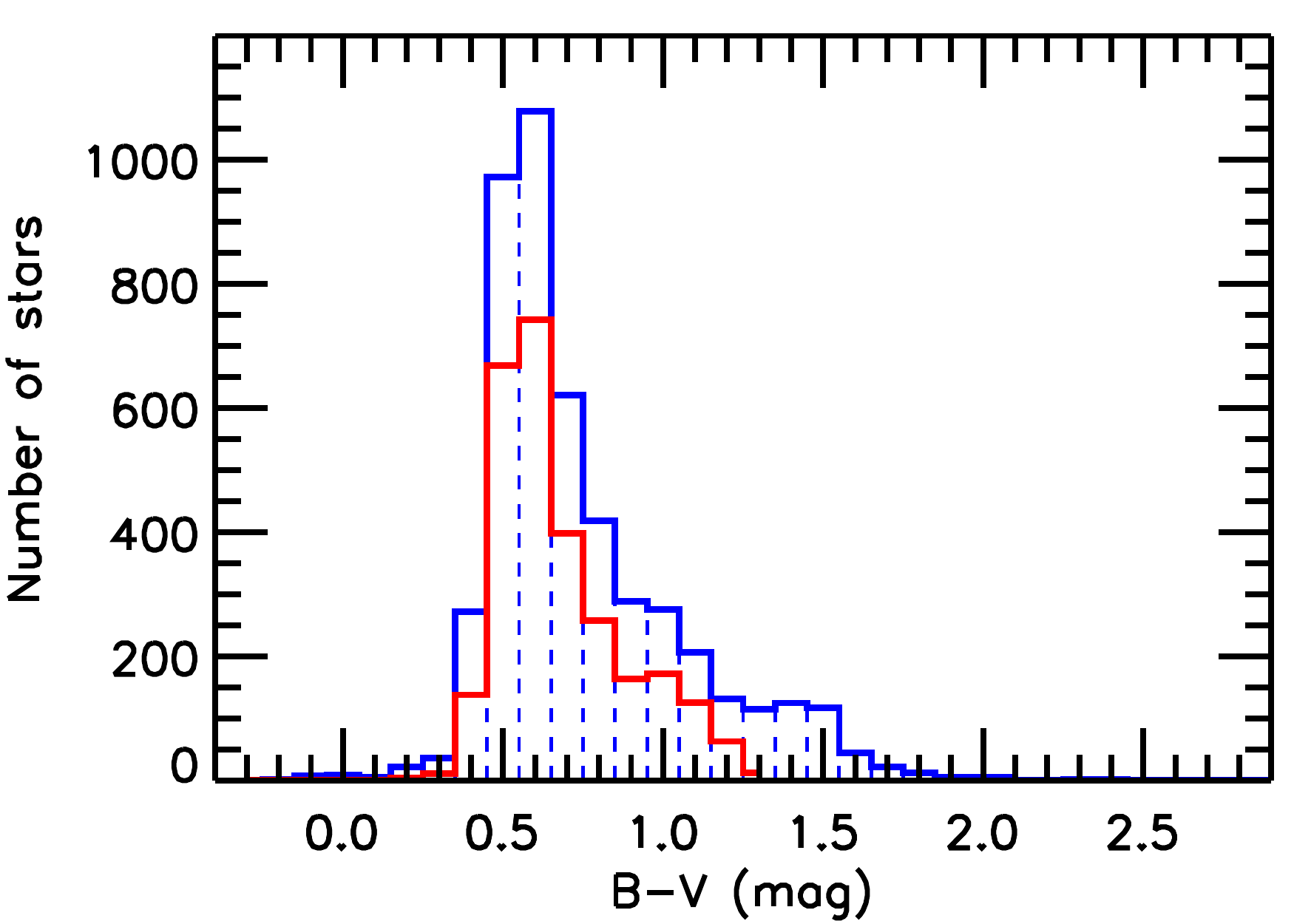}      
\caption{Histogram of $B-V$ colour for the whole compilation (in blue) and for the RV-STDs (in red).}
\label{f:histo_BV}
\end{center}
\end{figure}

Figure~\ref{f:histo_BV} shows the histograms of the $B-V$ colours for all the stars of the compilation, when available, with the corresponding histograms for the calibration stars.

All the stars not used for the calibrations were used as validation stars for the \RVS\ \citep{DR2-DPACP-54}. No constraint exists on their $V$ magnitude or spectral type, and their neighbourhood
was not inspected.  \\

\begin{figure*}[ht!]
\begin{center}
\includegraphics[width=14cm]{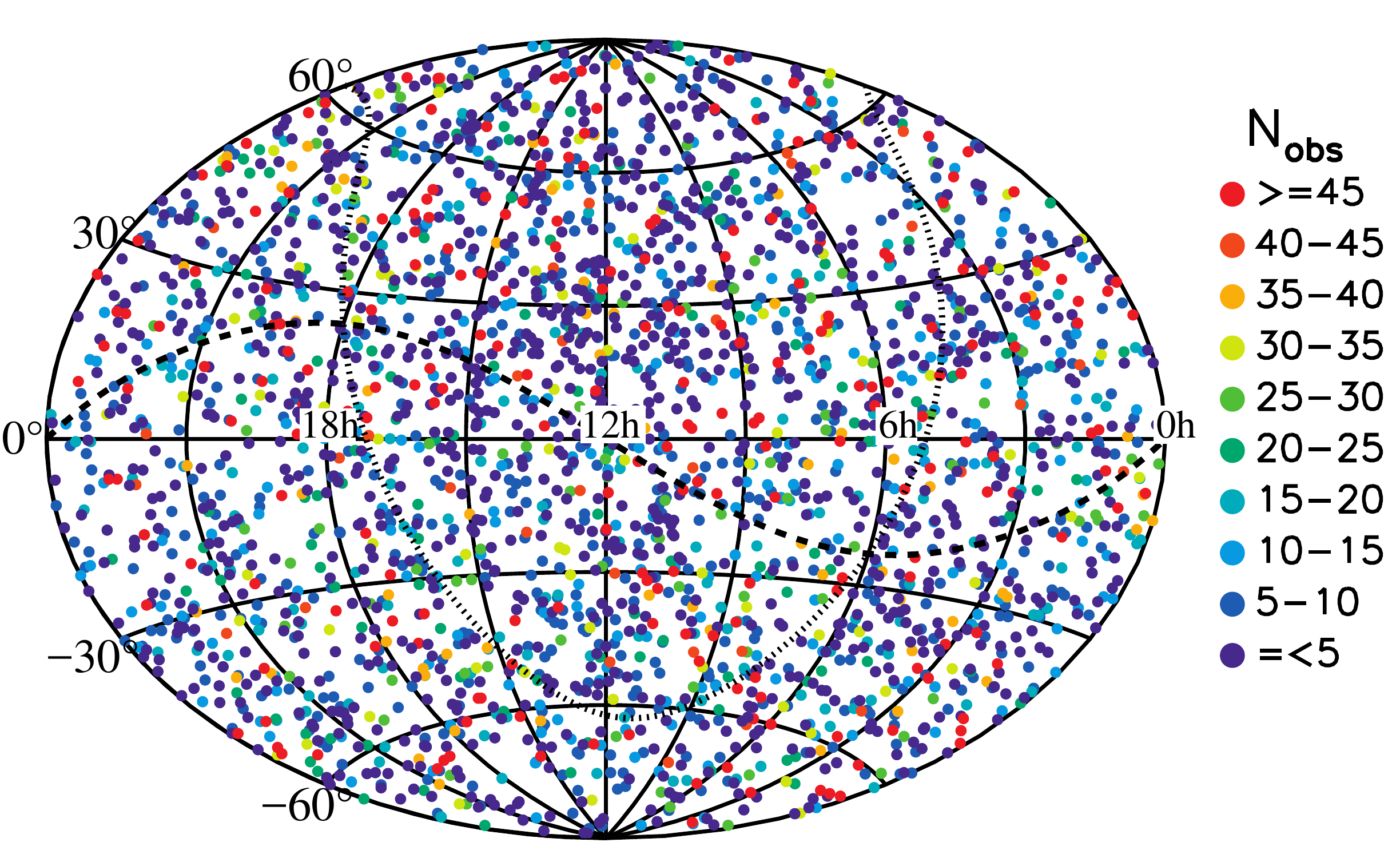}      
\caption{Distribution of the RV-STDs on the celestial sphere in equatorial coordinates. The dashed line indicates the projection of the Ecliptic plane, and the dotted line the projection of the Galactic plane. The colour code corresponds to the number of observations per star.}
\label{f:sky}
\end{center}
\end{figure*}

The  RV-STDs  used for the RV calibration in \gaia\ DR2 are shown on the celestial sphere in Fig.~\ref{f:sky}.  The homogeneous distribution on the sky is an important property that contributes to the proper \RVS\ calibration. \\

\begin{figure}[ht!]
\begin{center}
\includegraphics[width=\columnwidth]{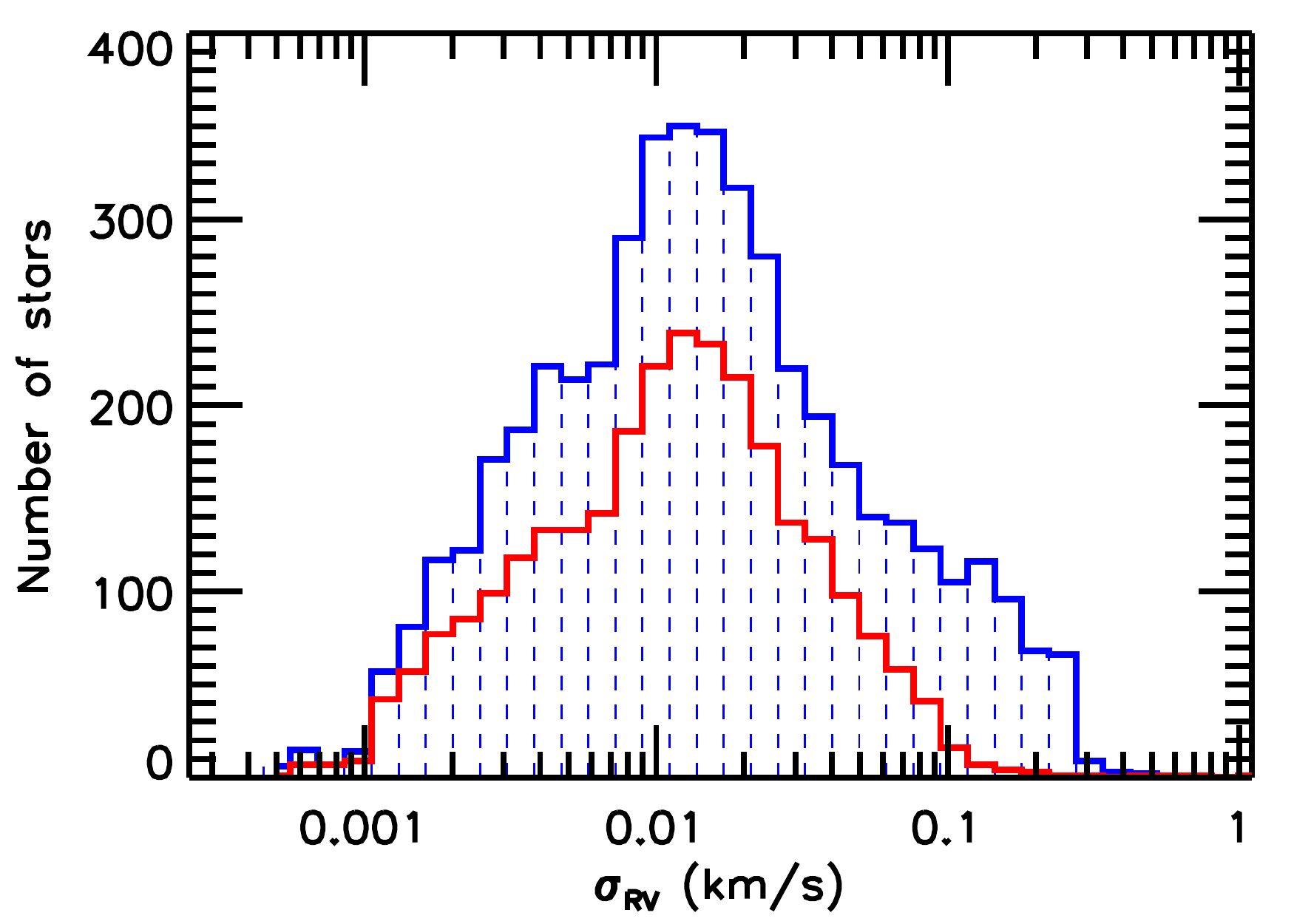}      
 \caption{Distribution of RV standard deviation, \srv, for the whole catalogue (in blue) and for the  RV-STDs (in red).}
\label{f:histo_var}
\end{center}
\end{figure}

Figure~\ref{f:histo_var} shows the histograms of the RV standard deviation, \srv, for the whole catalogue and for the RV-STDs.  The RV-STDs are highly stable, with a median \srv\ of 13 \ms\ and a 90\% percentile of 44 \ms. A few validation stars with \srv $>$ 0.3 \kms\ are included despite the cut made at this value. They correspond to RV-STDs  used for \gdrtwo\ that were discovered to be variable afterwards, as mentioned above.

\section{Comparison with other catalogues}
\label{s:comp}
\begin{figure}[ht!]
\begin{center}
\includegraphics[width=\columnwidth]{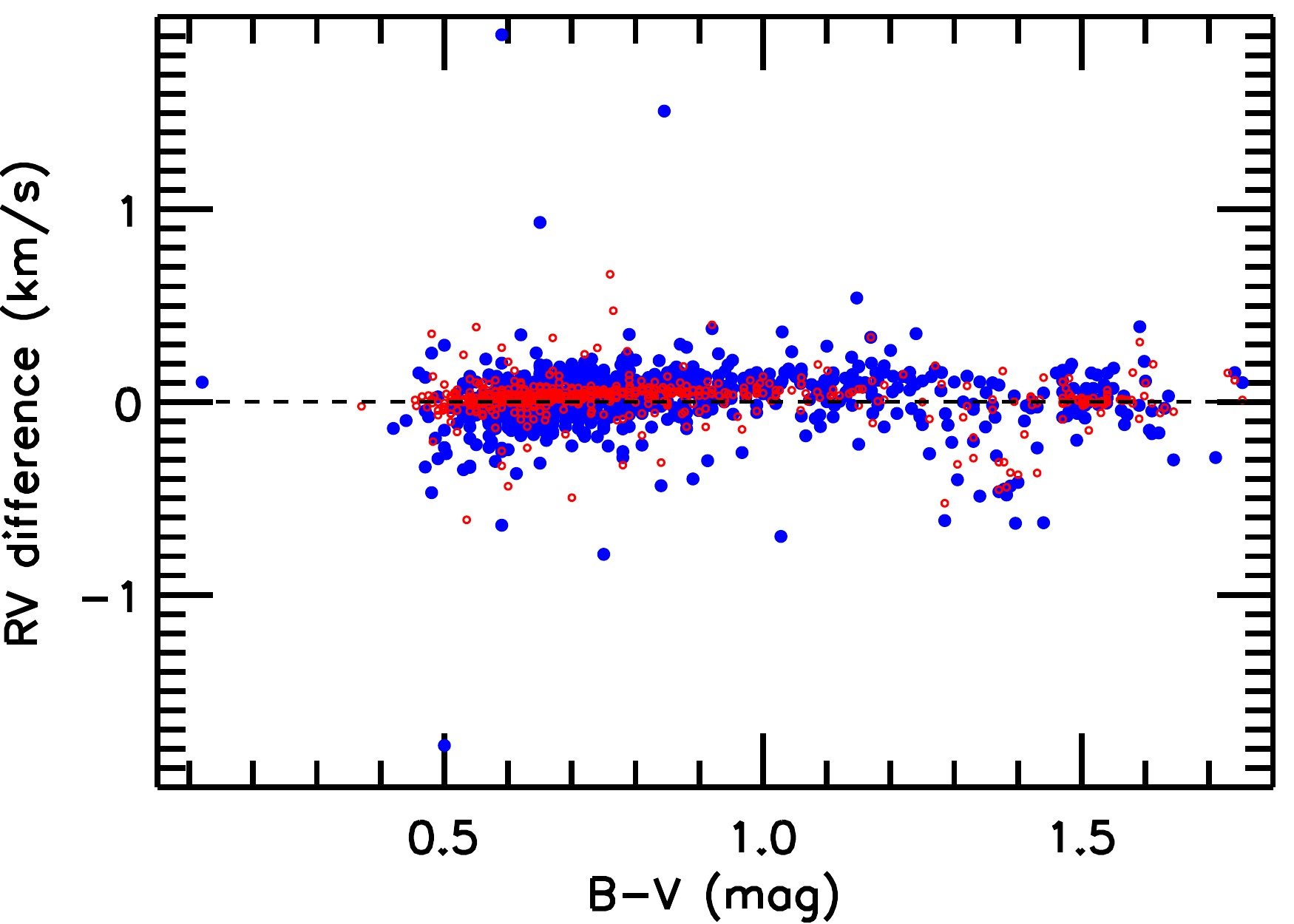}      
 \caption{RV difference between our measurements and those of \cite{nid02} (in red) and \cite{chu12} (in blue).}
\label{f:chu-nid}
\end{center}
\end{figure}

Each RV catalogue has its own RV ZP attached to the system and to the software used for the data processing.  In particular, the use of masks and/or specific synthetic spectra of different stellar atmosphere models can be different from one system to the next, as mentioned previously. 
Here we compare our catalogue with other large catalogues of ground-based RVs with typical precisions of the order of or better than 1 \kms. The results are presented in Table~\ref{t:comp}. Our results agree excellently well with the catalogues of \cite{nid02} and \cite{chu12}, as shown also in Fig.~\ref{f:chu-nid}. By construction, these two catalogues have the same velocity ZP, defined  by observations of the day sky and the asteroid Vesta. Thus it is expected that our ZPs coincide very well, with an offset as small as  26 \ms. The low dispersion obtained for the comparison, 26 \ms\ for \cite{nid02} and 72 \ms\ for \cite{chu12}, reflects the high internal precision of the compared catalogues.  \cite{nid02} quote an internal precision of 30 \ms\ , which suggests that their internal dispersion and our  internal dispersion are even lower (assuming that the dispersion of the RV difference combines the internal errors of the two compared catalogues in quadrature). This is consistent with our internal precision being 15 \ms\ (median \srv). Several stars with 1.2 $\leq B-V \leq$ 1.45 show systematic differences around $-300$ \ms. They mainly correspond to stars analysed with a K mask in our catalogue but with an M mask by \cite{nid02} and \cite{chu12}. This again indicates the difficulty of measuring the RV of M stars accurately. However, for stars with $B-V \geq$ 1.45, the ZPs are consistent with the remaining stars, showing that the M template used by \cite{nid02} and \cite{chu12} is consistent with the M5 mask used in the \sophie\ pipeline, and that the colour corrections applied to \elodie\ and \harps\ measurements are valid.

\begin{figure}[ht!]
\begin{center}
\includegraphics[width=\columnwidth]{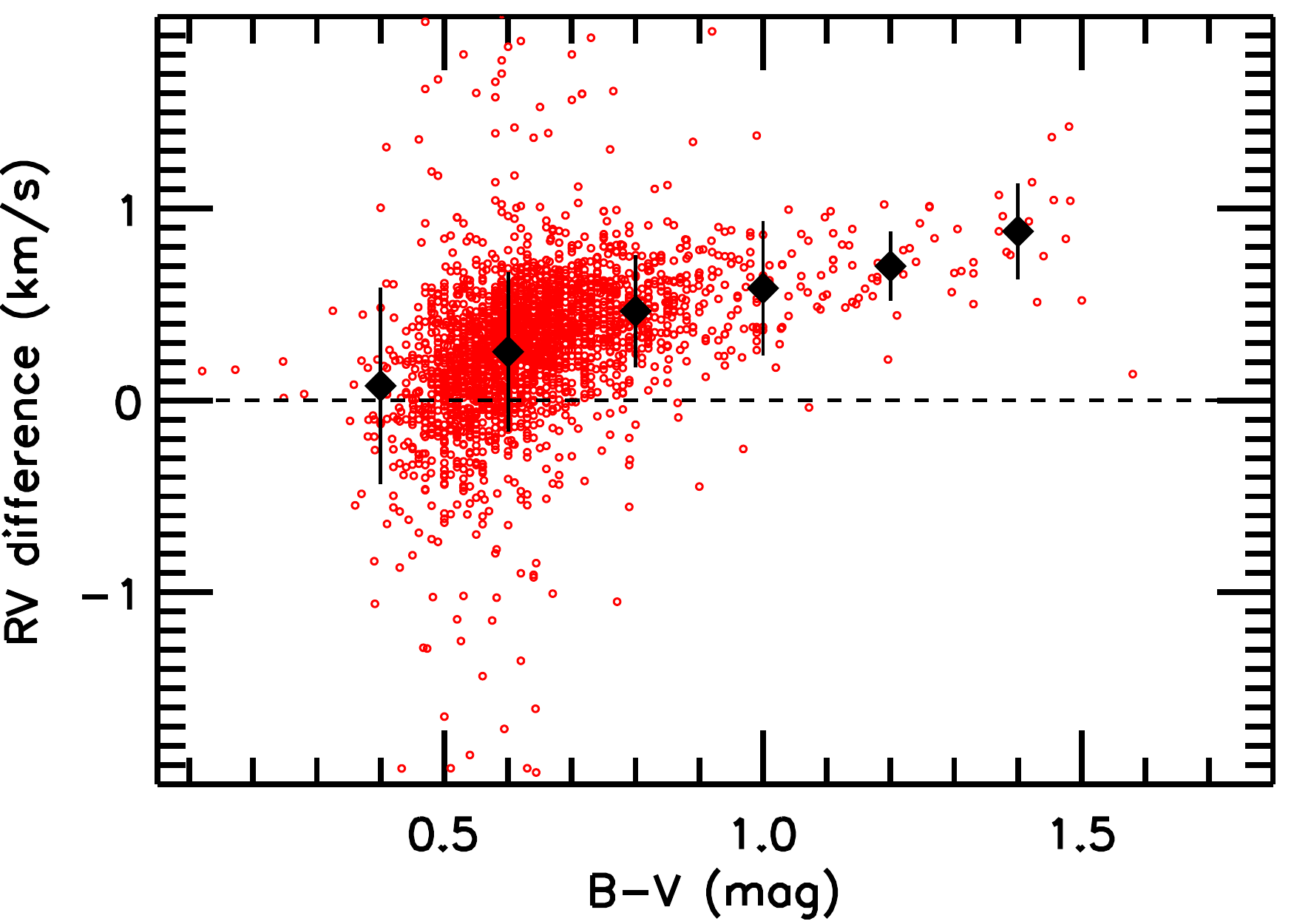}      
 \caption{RV difference between our measurements and those of the Geneva-Copenhagen survey \citep{nor04}.}
\label{f:nor}
\end{center}
\end{figure}

   About half of our sample is part of the Geneva-Copenhagen survey \citep{nor04}. The comparison shows a large dispersion and a clear trend with colour that reaches 1 \kms\ for the cooler stars (Fig.~\ref{f:nor}). No colour trend is observed with RAVE DR5 \citep{kun17}, but the dispersion is higher, consistent with the internal precision of that survey being of the order of 1 \kms.

\begin{table}[h]
 \centering
 \caption{Comparison of our catalogue with other ground-based RV catalogues, NID \citep{nid02}, CHU \citep{chu12}, WOR \citep{wor12}, FAM \citep{fam05}, MER \citep{mer08,mer09}, NOR \citep{nor04}, and RAV \citep[RAVE DR5][]{kun17}.
Median and median absolute deviation (MAD) are relative to the RV difference of our catalogue minus the other ones. $N_*$ and $N_{\rm o}$ refer to the number of stars in common and the number of outliers with RV differences larger than 3 \kms.}
 \label{t:comp}
\begin{tabular}{l|r|r|r|r}
\hline
Catalogue            &  $N_*$ & Median & MAD & $N_{\rm o}$ \\
            &        &  \ms   & \ms \\
\hline
NID & 567     &  26   & 26  & 0\\
CHU & 933     &  28   & 72 & 3\\
WOR & 280     &  $-$275   & 94   &  4 \\
FAM &  213     &  116   & 139  & 5\\
MER & 64     &  -5   & 203  & 2 \\
NOR &  2\,651     & 317   & 200   & 39\\
RAV & 315 & 292 & 960 & 43 \\
\hline
\end{tabular}
\end{table}

\section{\gaia\  RV measurements}
\label{s:rvs}
\begin{figure}[ht!]
\begin{center}
\includegraphics[width=\columnwidth]{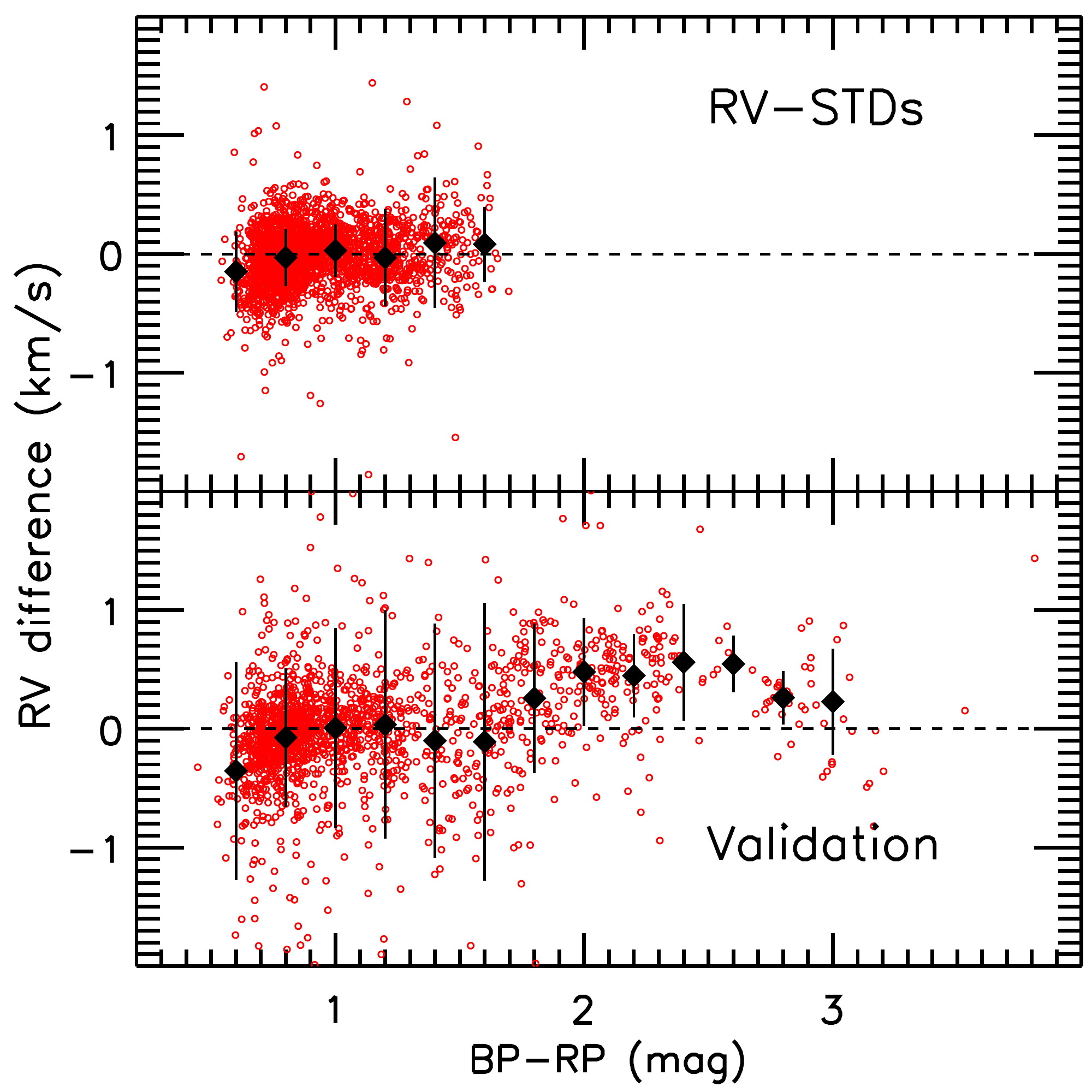}      
 \caption{RV difference (our catalogue -- \RVS)  as a function of the $G_{BP}-G_{RP}$ colour for the RV-STDs and for the validation stars}
\label{f:diff_rvs}
\end{center}
\end{figure}

Most of the stars of our catalogue (4\,290) have an \RVS\ measurement in \gdrtwo. Not all of them are included because of filters \citep{DR2-DPACP-39}.
Figure~\ref{f:diff_rvs} shows their RV difference (our catalogue minus \RVS) as a function of $G_{BP}-G_{RP}$ colour for the calibration stars and for the validation stars  
( $G_{BP}$ and $G_{RP}$ are the magnitudes in the two \gaia\ blue and red photometers, respectively). The low dispersion is remarkable (MAD = 165 \ms). The two catalogues are well aligned in their ZP as resulting from the calibration, except for the reddest stars with $G_{BP}-G_{RP} \gtrsim$ 1.8, corresponding to M stars, which show an offset of $\sim$500 \ms. This possibly reveals that the M template used by the \RVS\ pipeline is not consistent with the template used in the \sophie\ pipeline. 

A few outliers, not shown in the plots, might be explained by an error in the ground-based measurements such as an incorrect pointing or an error in the \RVS\ data, although we cannot exclude the presence of some binaries in the sample.

Figure~\ref{f:stable_stars} also represents the RVS velocities together with the ground-based velocities for two RV-STDs, and
it highlights their good agreement. The comparison of our catalogue with the \gaia\ RVs is extensively commented on as part of the \RVS\ data validation in \cite{DR2-DPACP-54}.

\section{Conclusions}
 We have presented a unique dataset considering the number of stars (4\,813), the homogeneity and precision of the RV measurements, and their time baseline (median 2\,307 days). The high internal precision of the catalogue is shown in the \srv\ of individual stars (median 15 \ms) and confirmed by comparison to the accurate RVs of \cite{nid02}. Our RVs were set on a ZP defined by repeated observations of asteroids. The resulting RV scale differs by only 26 \ms\ from that of \cite{nid02}. Our catalogue includes 2\,712 stars RV-STDs that fulfill the \RVS\ requirements for calibrations.  Creating a list of RV-STDs was mandatory for \gaia,\ but we took advantage of this task to build a larger compilation of RVs that was also essential to validate the first \gaia\ RVs published in DR2.

\begin{acknowledgements}
 This work has made use of results from the European Space Agency (ESA) space mission \gaia, the
data from which were processed by the \gaia\ Data Processing and Analysis Consortium (DPAC).
Funding for the DPAC has been provided by national institutions, in particular the institutions
participating in the \gaia\ Multilateral Agreement.
The \gaia\ mission website is
\url{http://www.cosmos.esa.int/gaia}. Many of the authors are current or past members of the ESA \gaia\
mission team and of the \gaia\ DPAC. We are indebted to the french "Action Sp\'ecifique \gaia" and  "Programme National Cosmologie et  Galaxies" for their financial support of the observing campaigns 
 and help in this project. We warmly thank Sergio Ilovaisky for helping us retrieve relevant data in OHP archives and for maintaining the nomenclature of the objects. 
Many thanks also go to the colleagues who made some observations for us. We also thank the staff 
maintaining the public archives of ready-to-use spectra at OHP, TBL and ESO. This research has made use of the SIMBAD and VizieR databases, operated at CDS, Strasbourg, France.  L.C acknowledges the financial support from the Comit\'e Mixto ESO-Chile and the DGI at University of Antofagasta. We acknowledge funding by the Belgian Federal Science Policy Office (BELSPO) through various Programme de D\'eveloppement d'Exp\'eriences Scientifiques (PRODEX) grants. TZ acknowledges funding from the Slovenian Research Agency (research core funding No. P1-0188). This publication makes use of data products from the Two Micron All Sky Survey, which is a joint project of the University of Massachusetts and the Infrared Processing and Analysis Center/California Institute of Technology, funded by the National Aeronautics and Space Administration and the National Science Foundation.
\end{acknowledgements}

\bibliographystyle{aa}
\bibliography{RV18}

\begin{thebibliography}{39}
\expandafter\ifx\csname natexlab\endcsname\relax\def\natexlab#1{#1}\fi

\bibitem[{{Anderson} \& {Francis}(2012)}]{xhip}
{Anderson}, E. \& {Francis}, C. 2012, Astronomy Letters, 38, 331

\bibitem[{{Arenou} {et~al.}(2018){Arenou}, {Luri}, {Babusiaux}, {Fabricius},
  {Helmi}, {Muraveva}, {C. Robin}, {Spoto}, {Vallenari}, {Antoja}, {Leclerc},
  {Reyl\'e}, {Robichon}, {Shih}, {Cantat-Gaudin}, {Bossini}, {Ruiz-Dern},
  {Turon}, {Sordo}, {A. Walton}, {Blanco-Cuaresma}, {Barache}, {A. Breddels},
  {Costigan}, {Diakit\'e}, {Figueras}, {Heu}, {Jordi}, {Lallement}, {Lambert},
  {M. Marrese}, {Massari}, {Fabrizio}, {Moitinho}, {Romero-G\'omez}, {Soria},
  {Soubiran}, {Veljanoski}, {Pancino}, {Bragaglia}, {Spagna}, \&
  {Tanga}}]{DR2-DPACP-39}
{Arenou}, F., {Luri}, X., {Babusiaux}, C., {et~al.} 2018, \aap\ (special issue
  for Gaia DR2)

\bibitem[{{Auri{\`e}re}(2003)}]{aur03}
{Auri{\`e}re}, M. 2003, in EAS Publications Series, Vol.~9, EAS Publications
  Series, ed. J.~{Arnaud} \& N.~{Meunier}, 105

\bibitem[{{Baranne} {et~al.}(1996){Baranne}, {Queloz}, {Mayor}, {Adrianzyk},
  {Knispel}, {Kohler}, {Lacroix}, {Meunier}, {Rimbaud}, \& {Vin}}]{bar96}
{Baranne}, A., {Queloz}, D., {Mayor}, M., {et~al.} 1996, \aaps, 119, 373

\bibitem[{{Bonifacio} {et~al.}(2017){Bonifacio}, {Caffau}, {Ludwig}, {Steffen},
  {Castelli}, {Gallagher}, {Ku{\v c}inskas}, {Prakapavi{\v c}ius}, {Cayrel},
  {Freytag}, {Plez}, \& {Homeier}}]{bon17}
{Bonifacio}, P., {Caffau}, E., {Ludwig}, H.-G., {et~al.} 2017, ArXiv e-prints
  [\eprint[arXiv]{1712.00024}]

\bibitem[{{Chiavassa} {et~al.}(2018){Chiavassa}, {Casagrande}, {Collet},
  {Magic}, {Bigot}, {Th{\'e}venin}, \& {Asplund}}]{chi18}
{Chiavassa}, A., {Casagrande}, L., {Collet}, R., {et~al.} 2018, \aap, 611, A11

\bibitem[{{Chubak} {et~al.}(2012){Chubak}, {Marcy}, {Fischer}, {Howard},
  {Isaacson}, {Johnson}, \& {Wright}}]{chu12}
{Chubak}, C., {Marcy}, G., {Fischer}, D.~A., {et~al.} 2012, ArXiv e-prints
  [\eprint[arXiv]{1207.6212}]

\bibitem[{{Crifo} {et~al.}(2012){Crifo}, {Chemin}, {Mignard}, {Jasniewicz},
  {Soubiran}, {Katz}, {Sartoretti}, \& {Hestroffer}}]{cri12}
{Crifo}, F., {Chemin}, L., {Mignard}, F., {et~al.} 2012, in SF2A-2012:
  Proceedings of the Annual meeting of the French Society of Astronomy and
  Astrophysics, ed. S.~{Boissier}, P.~{de Laverny}, N.~{Nardetto}, R.~{Samadi},
  D.~{Valls-Gabaud}, \& H.~{Wozniak}, 67--69

\bibitem[{{Crifo} {et~al.}(2007){Crifo}, {Jasniewicz}, {Soubiran},
  {Hestroffer}, {Siebert}, {Guerrier}, {Katz}, {Th{\'e}venin}, \&
  {Turon}}]{cri07}
{Crifo}, F., {Jasniewicz}, G., {Soubiran}, C., {et~al.} 2007, in SF2A-2007:
  Proceedings of the Annual meeting of the French Society of Astronomy and
  Astrophysics, ed. J.~{Bouvier}, A.~{Chalabaev}, \& C.~{Charbonnel}, 459

\bibitem[{{Crifo} {et~al.}(2010){Crifo}, {Jasniewicz}, {Soubiran}, {Katz},
  {Siebert}, {Veltz}, \& {Udry}}]{cri10}
{Crifo}, F., {Jasniewicz}, G., {Soubiran}, C., {et~al.} 2010, \aap, 524, A10

\bibitem[{{Cropper} {et~al.}(2018){Cropper}, {Katz}, {Sartoretti}, {Prusti},
  {de Bruijne}, {Chassat}, {Charvet}, {Boyadijan}, {Perryman}, {Sarri}, {Gare},
  {Erdmann}, {Munari}, {Zwitter}, {Wilkinson}, {Arenou}, {Vallenari},
  {G\'{o}mez}, {Panuzzo}, {Seabroke}, {Allende Prieto}, {Benson}, {Marchal},
  {Huckle}, {Smith}, {Dolding}, {Weingrill}, {Viala}, {Blomme}, {Baker},
  {Boudreault}, {Crifo}, {Soubiran}, {Fr\'{e}mat}, {Jasniewicz}, {Guerrier},
  {Guy}, {Turon}, {Jean-Antoine}, {Th\'{e}venin}, \& {David}}]{DR2-DPACP-46}
{Cropper}, M., {Katz}, D., {Sartoretti}, P., {et~al.} 2018, \aap\ (special
  issue for Gaia DR2)

\bibitem[{{De Pascale} {et~al.}(2014){De Pascale}, {Worley}, {de Laverny},
  {Recio-Blanco}, {Hill}, \& {Bijaoui}}]{dep14}
{De Pascale}, M., {Worley}, C.~C., {de Laverny}, P., {et~al.} 2014, \aap, 570,
  A68

\bibitem[{{Dravins}(1999)}]{dra99}
{Dravins}, D. 1999, in Astronomical Society of the Pacific Conference Series,
  Vol. 185, IAU Colloq. 170: Precise Stellar Radial Velocities, ed. J.~B.
  {Hearnshaw} \& C.~D. {Scarfe}, 268

\bibitem[{{Famaey} {et~al.}(2005){Famaey}, {Jorissen}, {Luri}, {Mayor}, {Udry},
  {Dejonghe}, \& {Turon}}]{fam05}
{Famaey}, B., {Jorissen}, A., {Luri}, X., {et~al.} 2005, \aap, 430, 165

\bibitem[{{Gaia~Collaboration}
  {et~al.}(2018{\natexlab{a}}){Gaia~Collaboration}, {Brown}, {Vallenari},
  {Prusti}, {de Bruijne}, \& {et al.}}]{DR2-DPACP-36}
{Gaia~Collaboration}, {Brown}, A.~G.~A., {Vallenari}, A., {et~al.}
  2018{\natexlab{a}}, \aap\ (special issue for Gaia DR2)

\bibitem[{{Gaia Collaboration} {et~al.}(2016{\natexlab{a}}){Gaia
  Collaboration}, {Brown}, {Vallenari}, {Prusti}, {de Bruijne}, {Mignard},
  {Drimmel}, {Babusiaux}, {Bailer-Jones}, {Bastian}, \& et~al.}]{gdr1b}
{Gaia Collaboration}, {Brown}, A.~G.~A., {Vallenari}, A., {et~al.}
  2016{\natexlab{a}}, \aap, 595, A2

\bibitem[{{Gaia~Collaboration}
  {et~al.}(2018{\natexlab{b}}){Gaia~Collaboration}, {Katz}, {Antoja},
  {Romero-G\'{o}mez}, {Drimmel}, {Reyl\'{e}}, \& {et al.}}]{DR2-DPACP-33}
{Gaia~Collaboration}, {Katz}, D., {Antoja}, T., {et~al.} 2018{\natexlab{b}},
  \aap\ (special issue for Gaia DR2)

\bibitem[{{Gaia Collaboration} {et~al.}(2016{\natexlab{b}}){Gaia
  Collaboration}, {Prusti}, {de Bruijne}, {Brown}, {Vallenari}, {Babusiaux},
  {Bailer-Jones}, {Bastian}, {Biermann}, {Evans}, \& et~al.}]{gdr1a}
{Gaia Collaboration}, {Prusti}, T., {de Bruijne}, J.~H.~J., {et~al.}
  2016{\natexlab{b}}, \aap, 595, A1

\bibitem[{{Katz} {et~al.}(2004){Katz}, {Munari}, {Cropper}, {Zwitter},
  {Th{\'e}venin}, {David}, {Viala}, {Crifo}, {Gomboc}, {Royer}, {Arenou},
  {Marrese}, {Sordo}, {Wilkinson}, {Vallenari}, {Turon}, {Helmi}, {Bono},
  {Perryman}, {G{\'o}mez}, {Tomasella}, {Boschi}, {Morin}, {Haywood},
  {Soubiran}, {Castelli}, {Bijaoui}, {Bertelli}, {Prsa}, {Mignot}, {Sellier},
  {Baylac}, {Lebreton}, {Jauregi}, {Siviero}, {Bingham}, {Chemla}, {Coker},
  {Dibbens}, {Hancock}, {Holland}, {Horville}, {Huet}, {Laporte}, {Melse},
  {Say{\`e}de}, {Stevenson}, {Vola}, {Walton}, \& {Winter}}]{kat04}
{Katz}, D., {Munari}, U., {Cropper}, M., {et~al.} 2004, \mnras, 354, 1223

\bibitem[{{Katz} {et~al.}(2018){Katz}, {Sartoretti}, {Cropper}, {Panuzzo},
  {Seabroke}, {Viala}, {Benson}, {Blomme}, {Jasniewicz}, {Jean-Antoine},
  {Huckle}, {Smith}, {Baker}, {Crifo}, {Damerdji}, {David}, {Dolding},
  {Fr\'{e}mat}, {Gosset}, {Guerrier}, {Guy}, {Haigron}, {Jan{\ss}en},
  {Marchal}, {Plum}, {Soubiran}, {Th\'{e}venin}, {Ajaj}, {Allende Prieto},
  {Babusiaux}, {Boudreault}, {Chemin}, {Delle Luche}, {Fabre}, {Gueguen},
  {Hambly}, {Lasne}, {Meynadier}, {Pailler}, {Panem}, {Royer}, {Tauran},
  {Zurbach}, {Zwitter}, {Arenou}, {Bossini}, {Gomez}, {Lemaitre}, {Leclerc},
  {Morel}, {Munari}, {Turon}, {Vallenari}, \& {\v{Z}erjal}}]{DR2-DPACP-54}
{Katz}, D., {Sartoretti}, P., {Cropper}, M., {et~al.} 2018, \aap\ (special
  issue for Gaia DR2)

\bibitem[{{Kunder} {et~al.}(2017){Kunder}, {Kordopatis}, {Steinmetz},
  {Zwitter}, {McMillan}, {Casagrande}, {Enke}, {Wojno}, {Valentini},
  {Chiappini}, {Matijevi{\v c}}, {Siviero}, {de Laverny}, {Recio-Blanco},
  {Bijaoui}, {Wyse}, {Binney}, {Grebel}, {Helmi}, {Jofre}, {Antoja}, {Gilmore},
  {Siebert}, {Famaey}, {Bienaym{\'e}}, {Gibson}, {Freeman}, {Navarro},
  {Munari}, {Seabroke}, {Anguiano}, {{\v Z}erjal}, {Minchev}, {Reid},
  {Bland-Hawthorn}, {Kos}, {Sharma}, {Watson}, {Parker}, {Scholz}, {Burton},
  {Cass}, {Hartley}, {Fiegert}, {Stupar}, {Ritter}, {Hawkins}, {Gerhard},
  {Chaplin}, {Davies}, {Elsworth}, {Lund}, {Miglio}, \& {Mosser}}]{kun17}
{Kunder}, A., {Kordopatis}, G., {Steinmetz}, M., {et~al.} 2017, \aj, 153, 75

\bibitem[{{Lindegren} \& {Dravins}(2003)}]{lin03}
{Lindegren}, L. \& {Dravins}, D. 2003, \aap, 401, 1185

\bibitem[{{Lopresto}(2000)}]{lop00}
{Lopresto}, J. 2000, in The Kth Reunion, ed. A.~G.~D. {Philip}, Vol.~18, 139

\bibitem[{{Marcy} {et~al.}(1987){Marcy}, {Lindsay}, \& {Wilson}}]{mar87}
{Marcy}, G.~W., {Lindsay}, V., \& {Wilson}, K. 1987, \pasp, 99, 490

\bibitem[{{Mayor} {et~al.}(2003){Mayor}, {Pepe}, {Queloz}, {Bouchy},
  {Rupprecht}, {Lo Curto}, {Avila}, {Benz}, {Bertaux}, {Bonfils}, {Dall},
  {Dekker}, {Delabre}, {Eckert}, {Fleury}, {Gilliotte}, {Gojak}, {Guzman},
  {Kohler}, {Lizon}, {Longinotti}, {Lovis}, {Megevand}, {Pasquini}, {Reyes},
  {Sivan}, {Sosnowska}, {Soto}, {Udry}, {van Kesteren}, {Weber}, \&
  {Weilenmann}}]{may03}
{Mayor}, M., {Pepe}, F., {Queloz}, D., {et~al.} 2003, The Messenger, 114, 20

\bibitem[{{Mermilliod} {et~al.}(2008){Mermilliod}, {Mayor}, \& {Udry}}]{mer08}
{Mermilliod}, J.~C., {Mayor}, M., \& {Udry}, S. 2008, \aap, 485, 303

\bibitem[{{Mermilliod} {et~al.}(2009){Mermilliod}, {Mayor}, \& {Udry}}]{mer09}
{Mermilliod}, J.-C., {Mayor}, M., \& {Udry}, S. 2009, \aap, 498, 949

\bibitem[{{Moultaka} {et~al.}(2004){Moultaka}, {Ilovaisky}, {Prugniel}, \&
  {Soubiran}}]{mou04}
{Moultaka}, J., {Ilovaisky}, S.~A., {Prugniel}, P., \& {Soubiran}, C. 2004,
  PASP, 116, 693

\bibitem[{{Nidever} {et~al.}(2002){Nidever}, {Marcy}, {Butler}, {Fischer}, \&
  {Vogt}}]{nid02}
{Nidever}, D.~L., {Marcy}, G.~W., {Butler}, R.~P., {Fischer}, D.~A., \& {Vogt},
  S.~S. 2002, ApJS, 141, 503

\bibitem[{{Nordstr{\"o}m} {et~al.}(2004){Nordstr{\"o}m}, {Mayor}, {Andersen},
  {Holmberg}, {Pont}, {J{\o}rgensen}, {Olsen}, {Udry}, \& {Mowlavi}}]{nor04}
{Nordstr{\"o}m}, B., {Mayor}, M., {Andersen}, J., {et~al.} 2004, \aap, 418, 989

\bibitem[{{Pasquini} {et~al.}(2012){Pasquini}, {Brucalassi}, {Ruiz},
  {Bonifacio}, {Lovis}, {Saglia}, {Melo}, {Biazzo}, {Randich}, \&
  {Bedin}}]{pas12}
{Pasquini}, L., {Brucalassi}, A., {Ruiz}, M.~T., {et~al.} 2012, \aap, 545, A139

\bibitem[{{Perruchot} {et~al.}(2008){Perruchot}, {Kohler}, {Bouchy}, {Richaud},
  {Richaud}, {Moreaux}, {Merzougui}, {Sottile}, {Hill}, {Knispel}, {Regal},
  {Meunier}, {Ilovaisky}, {Le Coroller}, {Gillet}, {Schmitt}, {Pepe}, {Fleury},
  {Sosnowska}, {Vors}, {M{\'e}gevand}, {Blanc}, {Carol}, {Point}, {Laloge}, \&
  {Brunel}}]{per08}
{Perruchot}, S., {Kohler}, D., {Bouchy}, F., {et~al.} 2008, in \procspie, Vol.
  7014, Ground-based and Airborne Instrumentation for Astronomy II, 70140J

\bibitem[{{Queloz} {et~al.}(2001){Queloz}, {Mayor}, {Udry}, {Burnet},
  {Carrier}, {Eggenberger}, {Naef}, {Santos}, {Pepe}, {Rupprecht}, {Avila},
  {Baeza}, {Benz}, {Bertaux}, {Bouchy}, {Cavadore}, {Delabre}, {Eckert},
  {Fischer}, {Fleury}, {Gilliotte}, {Goyak}, {Guzman}, {Kohler}, {Lacroix},
  {Lizon}, {Megevand}, {Sivan}, {Sosnowska}, \& {Weilenmann}}]{que01}
{Queloz}, D., {Mayor}, M., {Udry}, S., {et~al.} 2001, The Messenger, 105, 1

\bibitem[{{Sartoretti} {et~al.}(2018){Sartoretti}, {Katz}, {Cropper},
  {Panuzzo}, {Seabroke}, {Viala}, {Benson}, {Blomme}, {Jasniewicz},
  {Jean-Antoine}, {Huckle}, {Smith}, {Baker}, {Crifo}, {Damerdji}, {David},
  {Dolding}, {Fr\'{e}mat}, {Gosset}, {Guerrier}, {Guy}, {Haigron},
  {Jan{\ss}en}, {Marchal}, {Plum}, {Soubiran}, {Th\'{e}venin}, {Ajaj}, {Allende
  Prieto}, {Babusiaux}, {Boudreault}, {Chemin}, {Delle Luche}, {Fabre},
  {Gueguen}, {Hambly}, {Lasne}, {Meynadier}, {Pailler}, {Panem}, {Riclet},
  {Royer}, {Tauran}, {Zurbach}, {Zwitter}, {Arenou}, {Gomez}, {Lemaitre},
  {Leclerc}, {Morel}, {Munari}, {Turon}, \& {\v{Z}erjal}}]{DR2-DPACP-47}
{Sartoretti}, P., {Katz}, D., {Cropper}, M., {et~al.} 2018, \aap\ (special
  issue for Gaia DR2)

\bibitem[{{Soubiran} {et~al.}(2013){Soubiran}, {Jasniewicz}, {Chemin}, {Crifo},
  {Udry}, {Hestroffer}, \& {Katz}}]{sou13}
{Soubiran}, C., {Jasniewicz}, G., {Chemin}, L., {et~al.} 2013, \aap, 552, A64

\bibitem[{{Udry} {et~al.}(1999){Udry}, {Mayor}, \& {Queloz}}]{udr99}
{Udry}, S., {Mayor}, M., \& {Queloz}, D. 1999, in Astronomical Society of the
  Pacific Conference Series, Vol. 185, IAU Colloq. 170: Precise Stellar Radial
  Velocities, ed. J.~B. {Hearnshaw} \& C.~D. {Scarfe}, 367

\bibitem[{{Worley} {et~al.}(2012){Worley}, {de Laverny}, {Recio-Blanco},
  {Hill}, {Bijaoui}, \& {Ordenovic}}]{wor12}
{Worley}, C.~C., {de Laverny}, P., {Recio-Blanco}, A., {et~al.} 2012, \aap,
  542, A48

\bibitem[{{Zacharias} {et~al.}(2013){Zacharias}, {Finch}, {Girard}, {Henden},
  {Bartlett}, {Monet}, \& {Zacharias}}]{ucac4}
{Zacharias}, N., {Finch}, C.~T., {Girard}, T.~M., {et~al.} 2013, \aj, 145, 44

\bibitem[{{Zwitter} {et~al.}(2007){Zwitter}, {Mignard}, \& {Crifo}}]{zwi07}
{Zwitter}, T., {Mignard}, F., \& {Crifo}, F. 2007, \aap, 462, 795

\end{thebibliography}

\end{document}